\documentclass[a4paper]{revtex4}

\usepackage{amsmath}
\usepackage{graphicx}
\usepackage[final]{listings}

\begin{document}
    
\title{Metaprogramming Applied to Numerical Problems}


\author{Mario Mulansky}
\author{Karsten Ahnert}
\affiliation{Department of Physics and Astronomy, University of Potsdam}



\maketitle

Metaprogramming is a programming technique where a computer program is written that writes or manipulates another program (or itself) that then solves a given task.
This approach has several advantages over classical programming.
Firstly, it leads to more efficient programming because the process of generating code is automated as you basically write code that writes code.
Secondly, it might also lead to higher performance because using automatic code generation one can more easily produce code that can be optimized more efficiently by the compiler.
A popular example is the Matrix Template Library (MTL) where Template Metaprogramming is used to create a high performance linear algebra library~\cite{Siek-Lumsdaine-98}.

In C++ this is mostly realized by the use of templates and is then called \emph{Template Metaprogramming} \cite{Alexandrescu-01,Gamma-94}.
Templates are placholders for data types that are replaced -- \emph{instantiated} -- by the compiler at compile time.
Although templates where originally introduced to provide the functionality of parametrized types \cite{Stroustrup-89}, it was quickly realized that they are indeed much more powerful and form a programming language themself \cite{Veldhuizen-03}.
The first examples using C++ templates for Metaprogramming calculated prime numbers and were presented by E.~Unruh on the C++ standardization meeting in 1994 in San Diego \cite{Unruh-94}.
As most simple and instructive example serves the calculation of the factorial of a number at compile time \cite{Veldhuizen-95}.
Nowadays, Template Metaprogramming has found many applications, the most sophisticated are collected in the boost libraries, like Boost.MSM, Boost.Units, Boost.Accumulators and many others.
In this work, we will use the ideas and techniques of Metaprogramming to implement numerical algorithms.
This will heavily rely on existing Metaprogramming libraries, namely Boost.MPL \cite{Abrahams-Gurtovoy-04} and Boost.Fusion \cite{boost-fusion} -- compile time libraries that simplify the creation of Metaprogramming algorithms.

Metaprogramming is especially suitable to numerical algorithms where an approximate result is found by the successive application of a previously known number of stages where the type of computation at each state is similar.
Perfect examples for such algorithms are the explicit Runge-Kutta schemes \cite{Butcher-03} for finding the solution of Ordinary Differential Equations (ODEs).
In practice, solving ODEs numerically is usually done by finding the solution to an Initial Value Problem (IVP):
\begin{equation}
 \dot x(t) = f(x,t), \qquad x(t=0) = x_0.
\end{equation}
A Runge-Kutta scheme is defined by its number of stages $s$ and its integration parameters $c_1 \dots c_s$, \hspace{.5em}$a_{21}, a_{31} , a_{32} , \dots , a_{s s-1}$ and $b_1 \dots b_s$.
The approximate solution for $x_1 \approx x(h)$ starting at $x_0$ is then found by 
\begin{equation}
 x_1 = x_0 + h\sum_{i=1}^s b_i F_i \qquad \text{where} \qquad F_i = f( x_0 + h\sum_{j=1}^{i-1} a_{ij} F_j , h c_i ).
\end{equation}
Depending on the number of stages $s$ and the choice of parameters $a$, $b$, $c$, the approximate solution $x_1$ is exact up to some order $p$, which means $|x_1 - x(h)| \propto h^{p+1}$.
The parameters $a$, $b$ and $c$ define the so-called Butcher tableau and fully describe the specific Runge-Kutta scheme.
At each stage $i$ the following calculations have to be performed:
\begin{align} \label{eqn:rk_scheme}
 F_i &= f( y_{i} , h c_i ), \qquad y_{i+1} = x_0 + h\sum_{j=1}^{i} a_{i+1,j} F_j, \qquad i=1\dots s-1\quad \text{with}\quad y_1 = x_0 \\
 F_s &= f(y_{s} , h c_s ) , \qquad x_1 = x_0 + h\sum_{j=1}^{s} b_{j} F_j.
\end{align}

In this work we introduce a generic C++ implementation of such a Runge-Kutta scheme based on Template Metaprogramming.
Generic here means that our algorithms performs \emph{any} Runge-Kutta scheme for which the user provides the Butcher array.
Listing~\ref{lst:rk4_short} shows the general usage of our generic implementation using the classical 4th order Runge-Kutta scheme \cite{Kutta-1901} as an example.
This scheme is the most common one and can be found in any textbook on numerical methods for ODEs, e.g.~\cite{Press-92}.
We start with defining the type representing the state of the system.
As we here use the famous Lorenz system \cite{Lorenz-63} as an introductory example, we use an array of doubles with length 3 as \lstinline+state_type+.
Before coming to the Runge-Kutta algorithm, the rhs of the ODE has to be defined.
This is done by the function \lstinline+lorenz+ that calculates $f(x,t)$ for the Lorenz system writting the result into the variable \lstinline+dxdt+.
\begin{lstlisting}[float=t,caption={Classical 4th order Runge-Kutta scheme realized using the generic algorithm. Note that some details are omitted here for simplicity, e.g.\ the coefficient types.},label=lst:rk4_short]
typedef array< double , 3 > state_type;

void lorenz( const state_type &x , state_type &dxdt , double t )
{ //defines the rhs of the ODE x' = f(x,t), the result is written to dxdt
	dxdt[0] = sigma * ( x[1] - x[0] );
	dxdt[1] = R * x[0] - x[1] - x[0] * x[2];
	dxdt[2] = x[0]*x[1] - b * x[2];
}

int main() {

	// define the butcher array
	const array< double , 1 > a1 = {{ 0.5 }};
	const array< double , 2 > a2 = {{ 0.0 , 0.5 }};
	const array< double , 3 > a3 = {{ 0.0 , 0.0 , 1.0 }};

	const coef_a_type a = fusion::make_vector( a1 , a2 , a3 );
	const coef_b_type b = {{ 1.0/6.0 , 1.0/3.0 , 1.0/3.0 , 1.0/6.0 }};
	const coef_c_type c = {{ 0.0 , 0.5 , 0.5 , 1.0 }};

	// create the runge kutta scheme using the butcher array in a,b,c
	explicit_rk< state_type , 4 > rk4_stepper( a , b , c );

	// define the state x = (x,y,z) of the lorenz system
	state_type x = {{ 10.0 * rand() , 10.0 * rand() , 10.0 * rand() }};

	// make a number of timesteps
	for( size_t i=0 ; i<num_of_steps ; ++i, t+=dt )
		rk4_stepper.do_step( lorenz , x , t , dt );

	cout << x[0] << " , " << x[1] << " , " << x[2] << endl;
}
\end{lstlisting}
As a next step the parameters $a$, $b$, $c$ for the Runge-Kutta scheme have to be defined, which is also done in terms of arrays of doubles.
For simplicity, we here omit details on the types of the parameter arrays, like \lstinline+coef_a_type+, but they are defined within our library and can be easily obtained.
Finally, the Runge-Kutta stepper is created as an instance of the class \lstinline+explicit_rk< state_type , 4 >+.
Here, we tell the stepper as template parameters which state type to use (\lstinline+array<double,3>)+ and how many stages the scheme has (four).
The constructor then takes the parameters defined above and we are ready to perform the integration by successive calls of the \lstinline+rk4_stepper.do_step+ function.

As seen in this example, with our library it is very easy to implement arbitrary Runge-Kutta schemes just by providing the corresponding Butcher arrays.
The calculation details are hidden in the class \lstinline+explicit_rk+ which injects the given parameters into the algorithm and prevents the user from rewriting similar code when using different Runge-Kutta schemes.
Of course, such a generic implementation can be obtained by a run-time implementation as well where basically a \lstinline+for+--loop iterates through the stages.
However, in such a case one finds a worse performance than for a direct implementation of the Runge-Kutta scheme as it is shown in our performance analysis.
The reason is that in a run-time implementation the number of stages is unknown to the compiler which means that it has less possibilities for optimization.
%

The essential part of our algorithm is the class template \lstinline+stage+ representing one stage of the Runge-Kutta scheme, as shown in Listing~\ref{lst:stage}.
This class is meant to store the relevant information required to calculate one stage of the scheme, that is the number of this stage $i$, the parameter $c_i$ and the parameters $a_{ij}$, $b_j$ for the last stage respectively.
\begin{lstlisting}[float=t,caption=Definition of the fundamental stage class.,label=lst:stage]
template< class T , size_t i >
struct stage // general (intermediate) stage
{
	T m_c; // parameter c_i
	boost::array< T , i > m_a; // parameters a_i+1,1 ... a_i+1,i
							   // b_1 .. b_j respectively for the last stage
};
\end{lstlisting}
Suppose we have an instance of such a stage filled with appropriate parameter values, than one stage of the according Runge-Kutta scheme can be calculated from Eqs.~\eqref{eqn:rk_scheme}.
Listing~\ref{lst:calc_stage} shows the \lstinline+calc_stage+ function that does exactly this computation.
\begin{lstlisting}[float=t,caption=The \lstinline+calc_stage+ function performing one stage calculation.,label=lst:calc_stage]
state_type x; // the current state x
state_type x_tmp; // to store the intermediate result y_i+1
state_type F[stage_count]; // store the derivatives F_i
System system; // the rhs of the ODE, usually a function pointer

template< typename T , size_t i >
void calc_stage( const stage< T , i > &stage )
{  // performs one stage calculation of the i-th stage
	// obtain the rhs using x for the first stage, y_i otherwise
	if( i == 1 )
		system( x , F[i-1] , t + stage.c * dt );
	else
		system( x_tmp , F[i-1] , t + stage.c * dt );

	if( i < last_stage ) // performs the calculation x_tmp = x + dt * sum_j a_ij*F_j
		algebra<i>::vec_add( x_tmp , x , stage.a , F , dt);
	else // x = x + dt * sum_j b_j*F_j for the last, respectively
		algebra<i>::vec_add( x , x , stage.a , F , dt);
}
\end{lstlisting}
The actual calculation is hidden in the function \lstinline+vec_add+, which is implemented straight forward by a for-loop iterating through $j=1\dots i$.
It is precisely at this point of the code where the strength of our approach comes into play:
The stage number $i$ of the stage to be currently calculated is known at compile time which allows for an efficient optimization of the code.
Having a successive number of instances of the stage class with increasing stage number $i=1\dots s$ and filled with the parameters $c_i$, $a_{ij}$ and $b_i$ the complete Runge-Kutta step can be performed by successive calls of the \lstinline+calc-stage+-function.
The remaining task is to provide a convenient way to create and initialize the stages and to perform the steps.
Therefore, we implemented the \lstinline+explicit_rk+ class template which can be easily initialized and provides the \lstinline+do_step+ method for performing one complete Runge-Kutta step, as seen in Listing~\ref{lst:rk4_short}.
Here, we made heavy use of the Boost.MPL and Boost.Fusion libraries which allowed us to write a sophisticated Meta-Algorithm in about 200 lines of code.
Furthermore, the support of embedded Runge-Kutta methods which allow for error estimation is straight forward as it basically involves just the addition of another stage used to calculate the error estimate.
This is also already implemented in our library and can be used by instantiating the \lstinline+explicit_error_rk+ class template.

\begin{figure}[!t]
	\includegraphics[width=0.38\textwidth]{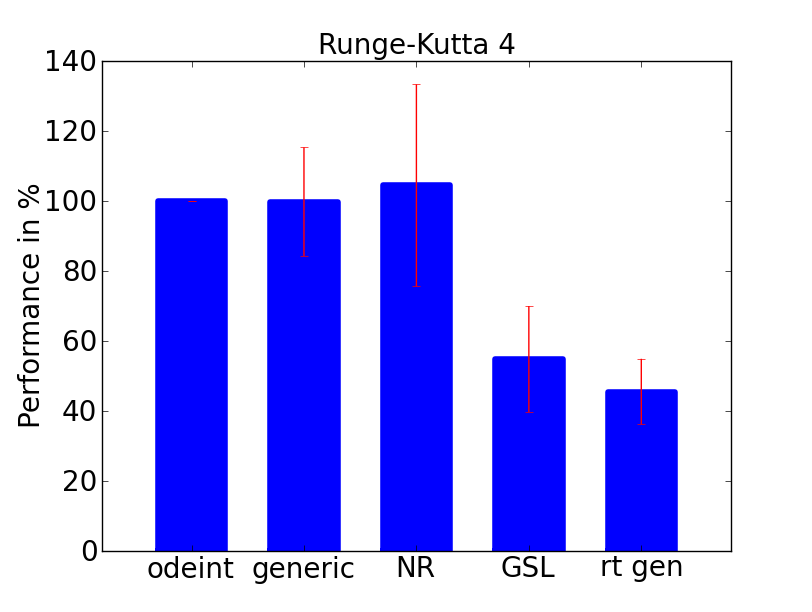} \hfill
	\includegraphics[width=0.38\textwidth]{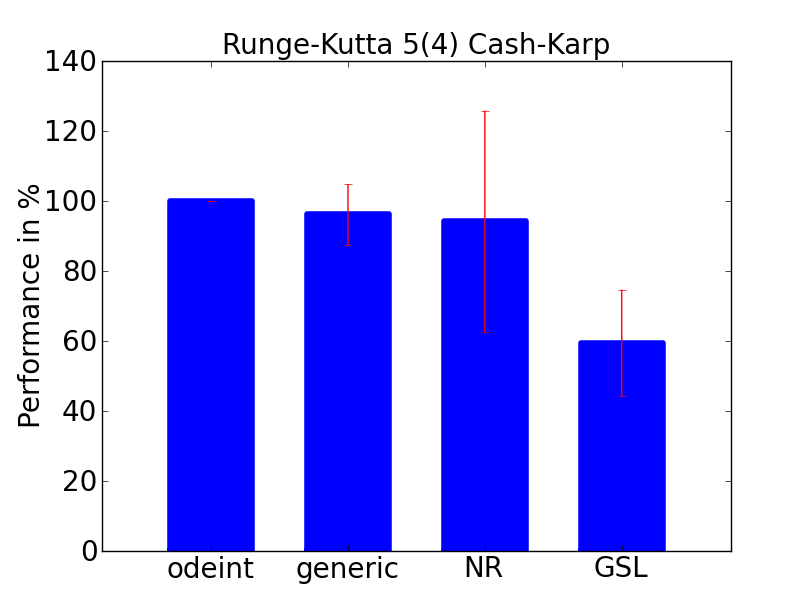} \hfill
\begin{footnotesize}
	\begin{tabular}[b]{c}
	\textbf{Processors:} \\
		Intel Core i7 830 \\
		Intel Core i7 930 \\
		Intel Xeon X5650 \\
		Intel Core2Quad Q9550 \\
		AMD Opteron 2224 \\
		AMD PhenomII X4 945 \\
		\hline
		\textbf{Compilers:} \\
		gcc 4.3 , 4.4 , 4.5 , 4.6 \\
		intel icc 11.1 , 12.0 \\
		msvc 9.0
	\end{tabular}
\end{footnotesize}
	\caption{Performance of different Runge-Kutta implementations. The left panel shows the results for the classical RK4 scheme, while on the right the RK5(4) Cash-Karp sheme was used. The graphs show the performance relative to the \texttt{odeint}-implementation averaged over several runs on different machines using different compilers (see table on the right).}
	\label{fig:perf}
\end{figure}

The crucial point for us was to create a generic library that shows similar performance to direct implementations of the Runge-Kutta schemes.
To check whether we have met this goal we did excessive performance tests comparing different implementations with the new generic approach.
As reference implementation we used the routines in \texttt{odeint} \cite{odeint} as they provide state-of-the art implementations of ODE solvers.
We also adapted the algorithms from the ``Numerical Recipes`` (NR) and made use of the routines from the Gnu Scientific Library (GSL).
Finally, we  also implemented a runtime version of a generic Runge-Kutta without any Template Metaprogramming (rt gen, RK4 only).
As actual integration schemes we first used the classical Runge-Kutta-4 algorithm, but also measured the runtime for the RK5(4) Cash-Karp scheme which involves error estimates.
The measurements were performed for a number of modern machines and with different compilers, see table in Fig.~\ref{fig:perf}.
For each run we computed the relative performance (one over runtime) based on the runtime of \texttt{odeint}.
The results are shown in Fig.~\ref{fig:perf}.
As seen there, \texttt{odeint}, our new generic implementation and the algorithms from the Numerical Recipes show basically the same performance, while the GSL and the runtime generic algorithm are significantly slower by about a factor 2.
It should be noted that we found heavy fluctuation up to 50\% between different compilers and in general the latest gcc version showed the best performance.
However, we conclude from these tests that our generic Runge-Kutta algorithm based on Template Metaprogramming is indeed competitive in terms of run-time to direct implementations.
This is a direct consequence of the compile-time programming as the run-time implementation is significantly slower.

As a summary, we here present a C++ implementation of a generic Runge-Kutta algorithm based on Template Metaprogramming.
With generic we mean that our algorithm runs \emph{any} explicit Runge-Kutta scheme where the user has only to provide the parameters $a$, $b$, $c$.
The use of Template Metaprogramming allows for a run-time performance of the Runge-Kutta schemes comparable to, sometimes even faster than, a direct, straight forward implementation of the scheme.
The advantage of our approach is that we completely separated the algorithm from the underlying parameters which reduces the effort of implementing new Runge-Kutta schemes to simple parameter initializations.

\bibliographystyle{unsrt}
\bibliography{books,articles}

\end{document}